\documentclass[aps,pra,showpacs,twocolumn,superscriptaddress]{revtex4-1}
\usepackage{amsmath}
\usepackage{amsfonts}
\usepackage{amssymb}
\usepackage{graphicx}
\usepackage{bbm}
\usepackage{natbib}
\usepackage[dvipdfm]{hyperref}

\begin{document}
\title{Pulse control of sudden transition for two qubits in XY spin baths and quantum phase transition}
\author{Da-Wei Luo}
\affiliation{Zhejiang Institute of Modern Physics and Department of Physics, Zhejiang University, Hangzhou 310027, People's Republic of China}
\author{Hai-Qing Lin}
\affiliation{Beijing Computational Science Research Center, Beijing 100084, People's Republic of China}
\author{Jing-Bo Xu}
\email{xujb@zju.edu.cn}
\affiliation{Zhejiang Institute of Modern Physics and Department of Physics, Zhejiang University, Hangzhou 310027, People's Republic of China}
\author{Dao-Xin Yao}
\affiliation{State Key Laboratory of Optoelectronic Materials and Technologies, School of Physics and Engineering, Sun Yat-sen University, Guangzhou 510275, People's Republic of China}
\affiliation{Department of Physics and Astronomy, University of Tennessee, Knoxville, TN 37996, USA}
\date{\today}

\begin{abstract}
We study the dynamics of two initially correlated qubits coupled to their own separate spin baths modeled by a XY spin chain and find the explicit expression of the quantum discord for the system. A sudden transition is found to exist between classical and quantum decoherence by choosing certain initial states. We show that the sudden transition happens near the critical point, which provides a new way to characterize the quantum phase transition. Furthermore, we propose a scheme to prolong the transition time of the quantum discord by applying the bang-bang pulses.
\end{abstract}

\maketitle

There has been increasing interest in the study of quantum correlations and quantum decoherence since the recent developments in quantum computation ~\cite{Werlang2009,Xu2010,Lanyon2008,Datta2008,Mazzola2010,natu}. In early studies, entanglement has been widely used as a special kind of quantum correlations in quantum information processing. However, some recent studies show that entanglement is not the only type of nonclassical correlations which can support the implementation of quantum information processing. It has been found that the quantum discord has an advantage over entanglement and may be responsible for the speed-up of a mix-state based quantum computation~\cite{Lanyon2008,Datta2008}. Quantum discord, introduced by Ollivier and Zurek~\cite{Ollivier2001}, is defined as the difference between the mutual information and classical correlation. It has been shown that the quantum discord can be completely unaffetced by certain decoherence environment during an initial time interval~\cite{Mazzola2010,pra_hql} which means that there is a sudden transition from classical to quantum decoherence. This phenomenon has been verified by the recent experiment~\cite{Xu2010}.

Theoretically, a deep understanding on the sudden transition is still called for and there are only few cases been used to study quantum discord and sudden transition. The spin-bath environment~\cite{Schlosshauer2005,Quan2006} may provide a way to study this interesting phenomenon. It has been pointed out that the entanglement or the Loschmidt Echo for a single spin rapidly decreases near the critical point~\cite{Quan2006,Sun2007}. The Loschmidt Echo was introduced in NMR experiments to describe the hypersensitivity of the time evolution to the environmental effects. Some interesting questions arise: Does the sudden transition from classical to quantum decoherence exist in the system of quantum critical environment? What is the relationship between the sudden transition and the quantum phase transition?

\begin{figure}
  \includegraphics[scale=.5]{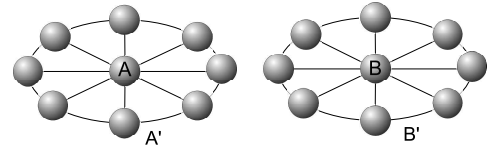}\\
  \caption{This is the schematic diagram of the system studied in this Letter. There is no interaction between qubits A and B. The spin baths are modeled by spin-XY interaction with periodic boundary conditions.}\label{spbth}
\end{figure}

In this Letter, we address these questions by studying the quantum discord dynamics of two initially correlated qubits coupled to their own spin baths modeled by spin-XY interaction. We obtain the explicit expression of the quantum discord dynamics and find that for a class of initial states, there exists a sudden transition~\cite{Mazzola2010} from classical to quantum decoherence in a finite time interval. It is shown that the transition time is sensitive to the initial state parameter, the coupling strength of the qubit-bath, the anisotropic coefficient of the XY spin bath and the external magnetic field. We also show that the sudden transition is closely associated with the quantum phase transition and only happens near the critical point for certain initial state and system setup parameters, which indicates the sudden transition may be used to describe the critical point of quantum phase transition. For example, it can be a new measure of the geometry of the ground state of a quantum many-body system like fidelity. Interestingly, the quantum discord for the two central qubits is completely unaffected by isotropic baths for this class of initial states. How to prolong the transition time of quantum discord in quantum information science is very important since it provides more time for the realization of quantum information and computation tasks. We propose a scheme to prolong the transition time by applying a train of bang-bang pulses~\cite{Rossini2008,Yang2008,Viola1998,epl}, which has been proven to be an effective tool to suppress decoherence induced by environment. Experimental suppression of polarization decoherence in a ring cavity using bang-bang pulses has been reported~\cite{prlbb}.


As an illustrative example, we consider two central qubits transversely coupled to their own spin baths modeled by the one dimensional anisotropic XY spin chain (See Fig.~\ref{spbth}). The Hamiltonian for the system is given by

\begin{eqnarray}
&H&=H_c+H_{b}+H_{int}, \nonumber \\
&H_c&=\tau^z_A+\tau^z_B, \nonumber\\
&H_b&=-J\sum_{l=1}^{N}\sum_{k=A',B'}\left(\frac{1+\gamma}{2}\sigma_{l,k}^x\sigma_{l+1,k}^x
+\frac{1-\gamma}{2}\sigma_{l,k}^y\sigma_{l+1,k}^y \right.\nonumber\\
&&\left.+\lambda \sigma_{l,k}^z\right), \nonumber\\
&H_{int}&=J \delta \sum_{l=1}^N(\tau^z_A\otimes\sigma_{l,A'}^z+\tau^z_B\otimes\sigma_{l,B'}^z),
\end{eqnarray}
where $\sigma^\alpha_i$ are Pauli matrices, $\tau^z=|e \rangle\langle e|-|g \rangle\langle g|$, $A',B'$ are the baths and $A,B$ the central qubits. We assume a periodic boundary condition for the chain environment, i.e. $\sigma_{N+1}^\alpha=\sigma_1^\alpha$, $\alpha=x,y,z$ and initially the two qubits are prepared in an X-type quantum state of the form $(\mathbbm{1}+\sum_{i=x,y,z}c_i \sigma_i \otimes \sigma_i)/4$, where $c_i\in[-1,1]$ are real numbers and $\mathbbm{1}$ is the identity matrix. We denote the initial state of each bath as $|\varphi_b(0) \rangle$. Tracing out the spin bath degree of freedom, we can obtain the reduced density matrix for the two central qubits as follows,
\begin{align}
	\rho_{AB}(t)&=\frac{1}{4}\{[(1+c_3)(|ee \rangle\langle ee|+|gg \rangle\langle gg|) \nonumber\\
	&+(1-c_3)(|eg \rangle\langle eg|+|ge \rangle\langle ge|) ]\nonumber\\
	&+[(c_1-c_2) R^2(t) |ee \rangle\langle gg|+(c_1+c_2)|R(t)|^2 |eg \rangle\langle ge| \nonumber\\
	&+h.c.]\},\label{dmt}
\end{align}
where $h.c.$ is the hermitian conjugate, $R(t)=\langle \varphi_-(t)|\varphi_+(t) \rangle$, $|\varphi_\pm(t) \rangle=e^{-i H_\pm t}|\varphi_b(0) \rangle$, and  $H_\pm$ is the corresponding effective Hamiltonian as derived in~\cite{Jordan1928,Lieb1961,Sachdev1999} and reads
\begin{equation}
	H_\pm=\left[\sum_{k}2\Lambda_k^{(\pm)} \left(b_{k,\pm} ^\dagger b_{k,\pm}-1/2\right)\right]\pm 1, \label{hb_d}
\end{equation}
where $\Lambda_k^{(\pm)}=-J\sqrt{\gamma^2\sin^2 k+(\cos k-h_\pm)^2}$, $h_\pm=\lambda \mp \delta$, $k=\frac{2\pi m}{L}$, and $m=-N/2 \ldots N/2-1$ for even chains. The baths are initially prepared in the ground state of $H_-$, $|\varphi_b(0) \rangle=|G \rangle_-$.

Using the Hamiltonian in Eq.~\ref{hb_d}, we derive the explicit expression of $R(t)$ as (taking $J=1$) $R(t)=e^{-i (\omega_+ - \omega_-)t} \prod_{k>0} \chi(k)$, where $\chi(k)=\cos^2 f_k+ \sin^2 f_k e^{-i4t \Lambda_{k}^{(+)}}$, $\omega_\pm=-\sum_k \Lambda_k^{(\pm)} \pm 1$, $f_k=\varphi_k^{(+)}-\varphi_k^{(-)}$ and $\tan 2\varphi_k^{(\pm)}=\frac{\gamma \sin k}{\cos k-h_\pm}$. The ground state energy level-crossing only happens at $\Lambda_k^{(\pm)}=0$ for some $k$, which is only possible if we take $\sin^2 k=0$ and $(\cos k-h_\pm)^2$ for $\gamma\neq 0$. When $\gamma=0$, the system reduces to the $XX$ spin model. Therefore, it can be easily shown that the quantum phase transition point is at $h_\pm=1$. It should be pointed out that the analytic expression we derived above is valid for general chain length $N$.

Next we investigate the dynamics of quantum discord for the two initially correlated qubits coupled to their own spin baths modeled by the XY interaction. For a two-qubit system, the total correlation can be measured by the mutual information $\mathcal{I}(\rho_{AB})=S(\rho_A)+S(\rho_B)-S(\rho_{AB})$, where $S(\rho)=-\text{tr}[\rho \log_2 \rho]$ is the von Neumann entropy. We choose~\cite{Mazzola2010} the initial state parameters $(c_1,c_2)=(1,-c_3)$. The reduced density matrices for each qubit is maximally mixed, i.e. $\rho_A=\rho_B=\mathbbm{1}/2$. Inserting this into the definition of mutual information leads to the explicit expression for $\mathcal{I}(\rho_{AB})$. Then the quantum discord is defined as
\begin{equation}
\mathcal{Q}=\mathcal{I}-\mathcal{C},
\end{equation}
where $\mathcal{C}$ is the classical correlation and can be expressed as~\cite{Ali2010}
\begin{eqnarray}
	\mathcal{C}(\rho_{AB})=\max_{\{B_k\}}[S(\rho_A)-S(\rho_{AB}|\{B_k\})]\label{cdef},
\end{eqnarray}
where $\{B_k\}$ is a complete set of projection operators performed on subsystem $B$. The quantum conditional entropy is measurement-based and is given by $S(\rho_{AB}|\{B_k\})=\sum_k p_k S(\rho_k)$, where $p_k$ corresponds to the probability of the $k$-th outcome of the measurement, and $\rho_k=Tr_B[(\mathbbm{1}\otimes B_k) \rho_{AB} (\mathbbm{1}\otimes B_k)]/p_k$, $p_k=\text{tr}[(\mathbbm{1}\otimes B_k) \rho_{AB} (\mathbbm{1}\otimes B_k)]$. In order to calculate the classical correlation, we need to choose a set of complete orthonormal projectors measuring the subsystem $B$. Using Eq.~\eqref{cdef}, the analytical expression for the classical correlation can be obtained. Then the quantum discord can now be written in a closed form
\begin{eqnarray}
    \mathcal{Q}&=&[(1-Y)\log_2(1-Y)+(1+Y)\log_2(1+Y)]/2,\nonumber\\
        &&Y=\min\{|R(t)|^2,c_3\}\label{qd_case}.
\end{eqnarray}

\begin{figure}
  \includegraphics[scale=.5]{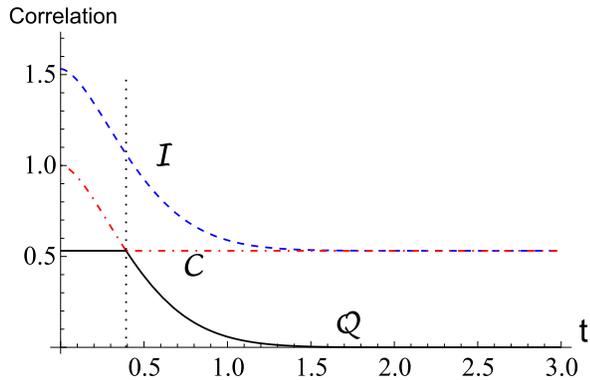}\\
  \caption{Time evolution of the mutual information (blue dashed line), classical correlation (red dash dotted line) and the quantum discord (black solid line) with $N=20$, $c_3=0.8$, $\lambda=0.9$, $\delta=0.3$ and $\gamma=0.4$. The vertical dotted line signifies the transition time $t_0$.}\label{sud0}
\end{figure}

From Eq.~\eqref{qd_case}, we can see that the system shows the sudden transition phenomenon once the reduced density matrix has the general form of Eq.~\eqref{dmt}, which does not depend on the details of $R(t)$. Generally speaking, local dephasing channels will give a reduced density matrix with the form of Eq.~\eqref{dmt}. 

In Fig.~\ref{sud0}, we plot the time evolution of the mutual information, classical correlation and the quantum discord with $N=20$, $c_3=0.8$, $\lambda=0.9$, $\delta=0.3$ and $\gamma=0.4$. It can be clearly seen that there exists a sudden transition from classical to quantum decoherence at time $t=t_0$. We want to note that this is the first time the sudden transition phenomenon is observed in spin environments.

\begin{figure}
  \includegraphics[scale=.5]{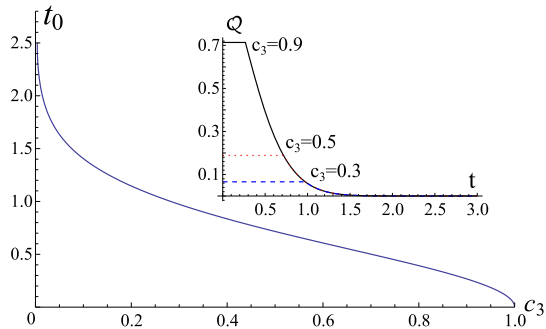}\\
  \caption{The sudden transition time $t_0$ is plotted as a function of the initial state parameter $c_3$. Inset: Time evolution of quantum discord for different values of $c_3$: $c_3=0.9$ (black solid line), $c_3=0.5$ (red dotted line), $c_3=0.3$ (blue dashed line). All other parameters are taken to be the same as Fig.~\ref{sud0}.}\label{stc3}
\end{figure}

The sudden transition time $t_0$ depends on the initial state parameter $c_3$ strongly, as shown in Fig.~\ref{stc3} with the inset showing the time evolution of quantum discord for some different values of $c_3$. We observe that for decreasing values of $c_3$, the quantum discord decreases while the transition time increases. After the sudden transition time $t_0$, they have the same decaying behavior.

\begin{figure}
  \includegraphics[scale=.5]{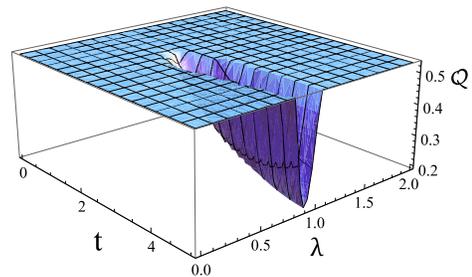}\\
  \caption{The quantum discord as a function of $\lambda$ and time $t$ is plotted with $c_3=0.8$, $\gamma=0.4$ and $\delta=0.05$. For the initial states chosen, the sudden transition only happens near the critical point $\lambda\sim1$, indicating the sudden transition is closely related to the quantum phase transition.}\label{stpt}
\end{figure}

Now we reveal the relationship between the sudden transition from classical to quantum decoherence and the quantum phase transition. As discussed previously, the quantum phase transition happens at $\lambda-\delta=1$. For some initial states, we find that the sudden transition between classical and quantum decoherence only occurs near the phase transition point, as is plotted in Fig.~\ref{stpt}, taking $c_3=0.8$, $\gamma=0.4$ and $\delta=0.05$, indicating that the sudden transition phenomenon exists in the system of quantum critical environment and only happens near the critical point for certain initial state and system setup parameters. Near the quantum critical point, the system has strong quantum fluctuations, which can cause $R(t)$ decay rapidly and the sudden transition may then happen. This means that the phenomenon of the sudden transition is closely associated with the quantum phase transition. The quantum discord can be a measure of the symmetry of the ground state of a quantum many-body system, like geometric tensor.

\begin{figure}
  \includegraphics[scale=.5]{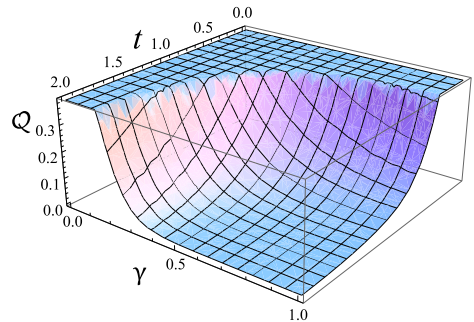}\\
  \caption{The quantum discord as a function of $\gamma$ and time $t$ is plotted with $c_3=0.7$, $\lambda=1$ and $\delta=0.25$. The sudden transition doesn't take place for $\gamma\rightarrow 0$, indicating the quantum discord of the two central qubits is completely unaffected by isotropic baths for this class of initial states.}\label{stg}
\end{figure}

In order to show the influence of the anisotropic coefficient of the spin bath on the transition time, we plot the quantum discord as a function of $\gamma$ and time $t$ in Fig.~\ref{stg}, from which we can see that the transition time varies gradually with $\gamma$, and as $\gamma$ approaches $0$, i.e. the environment approaches the isotropic spin ring, the transition time approaches infinity, indicating that for this case, the quantum discord for the two central qubits is completely frozen. The reason of this particular behavior is that when $\gamma=0$, the Bogoliubov rotation angle $\varphi^{(\pm)}=0$, therefore $|R(t)|^2=1$. Inserting this into the quantum discord for the qubits, Eq.~\eqref{qd_case}, we can see that the quantum discord now remains constant over time.


Finally, we propose a scheme to prolong the transition time for a system consisting of two initially correlated qubits coupled to their own spin baths modeled by spin-XY interaction by applying a train of bang-bang pulses, which has been proven to be an effective tool to suppress decoherence induced by environment. It is not difficult to obtain the time evolution operator of the pulses acting on the two central qubits simultaneously with a period of $2 \Delta t$ as $U_p=(i\sigma^x_A)\otimes(i\sigma^x_B)$. Denoting the evolution operator corresponding to the original Hamiltonian by $U_0(t)$, then the system evolution at time $t=2M \Delta t+\widetilde{t}$ is given by~\cite{Rossini2008}
\[
	U(t)=
	\begin{cases}
	U_0(\widetilde{t}) [U_\Delta]^M & \widetilde{t}<\Delta t\\
	U_0(\widetilde{t}-\Delta t)U_p U_0(\Delta t) [U_\Delta]^M & \widetilde{t}\geq \Delta t\\
	\end{cases}
\]
where $M$ is the integer part of $t/2\Delta t$, $\widetilde{t}$ the remainder, and $U_\Delta=U_p U_0(\Delta t)U_p U_0(\Delta t)$ for one cycle $2 \Delta t$. We focus on the periodic points only for simplicity. As pointed out in Ref.~\cite{Rossini2008}, this unitary time evolution corresponds to a monochromatic alternating magnetic field at resonance. After applying the pulses, the function $R(t)$ needs to be replaced by a new function $\mathcal{R}(t)$ and is given by
\begin{eqnarray}
    \mathcal{R}(t)&=&\langle G | (e^{i H_g \Delta t} e^{i H_e \Delta t})^M (e^{-i H_g \Delta t} e^{-i H_e \Delta t})^M|G \rangle \nonumber\\	
    &=&\langle G |\exp\{-M\Delta t^2[H_g,H_e]\}|G \rangle \nonumber\\
    &=&\prod_{k=1}^{N/2}\cos[4 \gamma\delta M\Delta t^2 \sin\frac{2\pi k}{N}],\label{bbqd}
\end{eqnarray}
at periodic points. The expression for the quantum discord remains of the same form.
\begin{figure}
  \includegraphics[scale=.5]{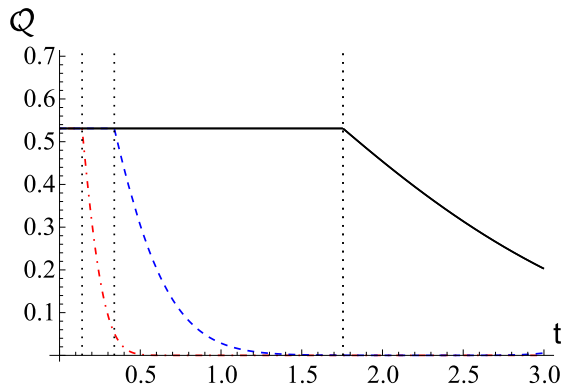}\\
  \caption{Time evolution of the quantum discord with $N=20$, $c_3=0.8$, $\delta=0.5$ and $\gamma=0.4$ is plotted for different values of $\lambda$: $\lambda=1$ (blue dashed line), $\lambda=0.1$ (red dash-dotted line) and with control pulses (black solid line) taking $\Delta t=0.3$. The vertical dotted line signifies the transition time.}\label{bb}
\end{figure}
In Fig.~\ref{bb}, we show the quantum discord without and with the pulse control. The blue dashed curve corresponds to the case where the coefficients are chosen to be near the phase transition point, which means the central qubits are coupled to a critical environment and the red dot-dashed line corresponds to a non-critical environment. It is quite clear that from Fig.~\ref{bb} that the bang-bang pulse is efficient to suppress the dephasing for both cases, prolonging the time over which the quantum discord does not decrease. Shorter pulses can extend the transition time even longer. It is interesting to experimentally realize this proposal to prolong the sudden transition time.


In conclusion, we have studied the quantum discord dynamics for two central qubits coupled to their own spin chain environment modeled by spin-XY interaction and found that for a class of initial states, there exists a sudden transition between classical and quantum decoherence, which is sensitive to the anisotropic coefficient, external magnetic field and initial state parameter. It is interesting to point out that the quantum discord for the two central qubits is completely unaffected by isotropic baths for this class of initial states. Moreover, the sudden transition phenomenon is found to be closely related to the quantum phase transition and only happens near the critical point for some initial states, which indicates the sudden transition may be used to describe the critical point of the quantum phase transition. Experimentally , the quantum dots can be fabricated by a superconductor or semiconductor and XY spin system can be realized by an array of Josephson junctions. They could be effectively coupled to each other. Finally, we proposed a scheme to prolong the transition time by applying a series of bang-bang pulses on the central qubits. These results may be used to improve the implementation of quantum tasks which are based on quantum correlations.

This project was supported by the National Natural Science Foundation of China (Grant No. 10774131 and 11074310) and China National Basic Research Program, Project No. 2011CB922200.



\end{document}